# Experimental and Theoretical Electronic Structure and Symmetry Effects in Ultrathin NbSe$_2$ Films

Cai-Zhi Xu[1,2,3], Xiaoxiong Wang[1,2,4], Peng Chen[1,2,3], David Flötotto[1,2], Joseph Andrew Hlevyack[1,2], Meng-Kai Lin[1,2], Guang Bian[5], Sung-Kwan Mo[3] and Tai-Chang Chiang[1,2,6]

[1]Department of Physics, University of Illinois at Urbana-Champaign, Urbana, Illinois 61801, USA

[2]Frederick Seitz Materials Research Laboratory, University of Illinois at Urbana-Champaign, Urbana, Illinois 61801, USA

[3]Advanced Light Source, Lawrence Berkeley National Laboratory, Berkeley, California 94720, USA

[4]College of Science, Nanjing University of Science and Technology, Nanjing 210094, China

[5]Department of Physics and Astronomy, University of Missouri, Columbia, Missouri 65211, USA

[6]Department of Physics, National Taiwan University, Taipei 10617, Taiwan

# Abstract

Layered quasi-two-dimensional transition metal dichalcogenides (TMDCs), which can be readily made in ultrathin films, offer excellent opportunities for studying how dimensionality affects electronic structure and physical properties. Among all TMDCs, $NbSe_2$ is of special interest; bulk $NbSe_2$ hosts a charge-density-wave phase at low temperatures and has the highest known superconducting transition temperature, and these properties can be substantially modified in the ultrathin film limit. Motivated by these effects, we report herein a study of few-layer $NbSe_2$ films, with a well-defined single-domain orientation, epitaxially grown on Gallium Arsenide (GaAs). Angle-resolved photoemission spectroscopy (ARPES) was used to determine the electronic band structure and the Fermi surface as a function of layer thickness; first-principles band structure calculations were performed for comparison. The results show interesting changes as the film thickness increases from a monolayer (ML) to several layers. The most notable changes occur between a ML and a bilayer, where the inversion symmetry in bulk $NbSe_2$ is preserved in the bilayer but not in the ML. The results illustrate some basic dimensional effects and provide a basis for further exploring and understanding the properties of $NbSe_2$.

Two-dimensional (2D) systems can host unique physical properties compared to their bulk counterparts; examples include the half-integer quantum Hall effect in graphene [1] and strong enhancement of photoluminescence in monolayer (ML) $MoS_2$ [2,3]. The vast varieties of ML transition metal dichalcogenides (TMDCs) provide many opportunities for exploring novel electronic, optical, and topological properties possibly relevant to device applications; they are also excellent choices for basic studies in connection with collective phenomena such as charge-density-wave (CDW) formation and superconductivity at low temperatures [4–9]. The present study focuses on $NbSe_2$, which, in the bulk 2H form, undergoes an incommensurate CDW transition at 33 K and then becomes superconducting at 7.2 K [10–12], the highest superconducting transition temperature among all TMDCs. The mechanism of the CDW transition and its relationship with superconductivity has been under intensive investigation. Early studies suggested that the CDW order in bulk $NbSe_2$ was driven by Fermi surface nesting [10,13], or alternatively by saddle-point singularities [14], but in more recent studies, momentum-dependent electron-phonon coupling was proposed to be the driving force [15–20]. These studies also suggested that the CDW and superconductivity in this system are possibly connected through either a cooperative or competitive interaction.

For ultrathin films of $NbSe_2$, Novoselov *et al.* found that $NbSe_2$ transformed from a metal in the bulk to a semimetal in the ML [21]. Xi *et al.* found that the CDW transition temperature of $NbSe_2$ increased from 33 K in the bulk to 145 K in the ML, while the corresponding superconducting transition temperature decreased from 7.2 to 3.1 K [5,6]. They found that the in-plane upper critical field for superconducting ML $NbSe_2$ was more than six times the Pauli paramagnetic limit, which was explained as an indication of Ising superconductivity [6]. Underlying these dimension-driven effects are changes in the electronic structure as the material becomes ultrathin. While the electronic structure of bulk $NbSe_2$ has been studied in detail [13,15,19], systematic investigations in the few-layer ultrathin film limit are still lacking. Ugeda *et al.* performed a study of the electronic structure of ML $NbSe_2$ grown on bilayer-graphene-terminated SiC(0001), but their results were complicated by crystal domain misalignment and angular averaging [4]. In the present work, we report the growth of high-quality ultrathin $NbSe_2$ films with a single crystallographic orientation on the B face of GaAs(111) using molecular beam epitaxy (MBE). The single-domain orientation of the films, made possible with the particular substrate choice, facilitated detailed band structure and Fermi surface mapping using angle-resolved photoemission spectroscopy (ARPES). Complementary first-principles calculations were performed for a comparison with the experimental results.

Our experiment was performed at the Advanced Light Source, Lawrence Berkeley National Laboratory, using the HERS endstation at Beamline 10.0.1. The growth process of NbSe$_2$ films is schematically illustrated in Fig. 1(a). Reflection high-energy electron diffraction (RHEED) was used to monitor the surface structure at each step (Fig. 1(b)). Substrates of GaAs(111)B were cleaned by repeated cycles of Ar ion sputtering and annealing at 450 °C in a MBE chamber attached to the beamline. The substrate after cleaning was held at 500 °C under a Se flux from a Knudsen cell to terminate/passivate the surface with Se [22]. The RHEED patterns (Fig. 1(b)) indicate that the surface quality was improved after Se passivation as evidenced by sharper patterns with a lower background, and the Se passivation facilitates the growth of Se-based TMDCs [23]. Subsequently, NbSe$_2$ films were grown by co-deposition of Nb and Se with the substrate held at 580 °C. The resulting RHEED pattern after 1-ML NbSe$_2$ growth was very sharp as seen in the bottom panel of Fig. 1(b). The spacings of the RHEED streaks for the 1-ML NbSe$_2$ film are larger than the corresponding spacings for the clean GaAs substrate (Fig. 1(c)). The results indicate that the NbSe$_2$ film adopts its own natural bulklike lattice constant, which is incommensurate with the GaAs substrate lattice constant, but the film maintains the same in-plane crystallographic orientation as the substrate. This type of incommensurate parallel epitaxy is common for the growth of van der Waals bonded layered materials. Subsequent ARPES measurements were performed with the sample maintained at 40 K using a Scienta R4000 analyzer; the energy and angular resolutions were 20 meV and 0.1°, respectively. The angle-integrated spectrum in Fig. 1(d) was taken with 80-eV photons. The measured intensities of the $3d$ core levels of Ga, Nb, As, and Se are consistent with a single layer of NbSe$_2$ on top of a GaAs substrate.

Our first-principles calculations of the film electronic structure were performed using the local-density approximation (LDA) implemented in the abinit package [24,25]. Relativistic pseudopotential functions constructed by Hartwigsen, Goedecker, and Hutter (HGH) were employed [26]. Calculations based on the generalized-gradient approximation (GGA) [27] and the PBE pseudopotential [28] were also performed for comparison; the results are very similar and therefore not shown here. A plane wave basis set with a cutoff energy of 400 eV and a Monkhorst-Pack $k$-space mesh of 11x11x1 were used. The substrate was ignored in the calculation, and the vacuum gap in the periodic slab model was set to be 15 Å. The in-plane lattice constant was obtained from energy minimization; the result $a = 3.45$ Å agrees closely with the bulk value of 3.44 Å [29].

The crystal structure of NbSe$_2$ is shown schematically in Figs. 2(a) and 2(b). A ML of NbSe$_2$ (Fig. 2(a)) has a hexagonal structure and consists of one Nb atomic layer sandwiched between two Se atomic layers. Its Brillouin zone is shown in Fig. 2(c). The unit cell for bulk 2H-NbSe$_2$ consists of two layers stacked as shown in Fig. 2(b). The 2-ML structure has inversion symmetry, but it is not the case for a ML. Low-energy-electron-diffraction (LEED) patterns from a 1-ML film (Fig. 2(d)) show a hexagonal pattern superimposed on a somewhat smaller hexagonal pattern arising from the GaAs substrate. For the 2-ML sample (Fig. 2(e)), only the NbSe$_2$ pattern remains; the GaAs substrate pattern is strongly attenuated and not seen. The hexagonal patterns from NbSe$_2$ are well aligned with respect to the GaAs substrate, further confirming the parallel epitaxy configuration. The sharpness of the LEED patterns attests to a high film quality and a high degree of orientational order, which is important for band mapping. A prior study of ultrathin NbSe$_2$ films shows a blend of misaligned NbSe$_2$ domains leading to a mixed electronic structure [4]. Our samples do not suffer from this ambiguity.

ARPES maps for 1-ML NbSe$_2$ along $\overline{\Gamma K}$ and $\overline{\Gamma M}$ are shown in Figs. 3(a) and 3(b), respectively, together with corresponding second-derivative maps [30]. The calculated band structure is overlaid over the second-derivative maps for comparison. Not all theoretical bands are seen in the experiment because of cross section variations, but for those bands seen, the agreement is fairly good, although some differences are not unexpected for such calculations. Our experimental band structure is consistent with features in the previous report on NbSe$_2$ grown on bilayer graphene [4] but is not affected by domain angular averaging. The system is metallic or semimetallic [5,21] as the top valence band crosses the Fermi level along both $\overline{\Gamma K}$ and $\overline{\Gamma M}$. This band is primarily derived from the Nb $4d$ orbital based on calculations, and it is responsible for the superconductivity and CDW in ML NbSe$_2$. Detailed views of this band from ARPES and the second derivatives along both $\overline{\Gamma K}$ and $\overline{\Gamma M}$ are shown in Fig. 3(c) for 1-, 2-, 4-, and 6-ML films. The computed band structure is also shown. The 1-ML film is non-centrosymmetric; its bands are generally spin split by spin-orbit coupling. This splitting is evident in Fig. 3(a) for the calculated Nb $4d$ band along the $\overline{\Gamma K}$ direction. However, the same band is spin-degenerate along the $\overline{\Gamma M}$ direction because of the protection offered by a perpendicular mirror plane that passes through this direction. The splitting along the $\overline{\Gamma K}$ direction is quite small and not resolved in the experiment. Nevertheless, this splitting is important and relevant to Ising superconductivity [6,31,32].

As the film thickness increases from 1 to 2 MLs, the number of bands should double (one band from each layer), and similarly for $N$-ML films each band should develop into $N$ bands. Figure 3(c) shows that, experimentally, the band along the $\overline{\Gamma M}$ direction in the 1-ML film indeed splits into two well-separated bands in the 2-ML film, and the splitting can be related to interlayer coupling. The experimentally measured splitting is roughly twice the theoretical prediction. Along the $\overline{\Gamma K}$ direction, the band also shows a substantial experimental splitting, which is again about twice the theoretical value. Interestingly, the number of bands along this direction remains to be two in going from 1 to 2 MLs, without doubling. This seemingly strange behavior is explained by the restoration of the inversion symmetry in going from 1 to 2 MLs, which eliminates the spin-orbit splitting, and each of the two bands for the 2-ML film is actually doubly (spin) degenerate. With the spin degeneracy taking into account, the number of bands is actually four. The restoration of the spin degeneracy can impact Ising pairing and thus affect the in-plane upper critical field [6]. Likewise, each of the two split bands along $\overline{\Gamma M}$ for the bilayer is spin degenerate. For the 4- and 6-ML films, the bands multiply as the number of layers, with each band being spin degenerate, according to theory. The band multiplication can be understood in terms of quantum size effects [33]. Experimentally, the additional bands in the 4- and 6-ML films are not well resolved because of the close energy spacings.

The Fermi surfaces of 1-, 2-, and 4-ML$NbSe_2$ films determined from ARPES are shown in Fig. 4(a); the corresponding second-derivative maps are shown in Fig. 4(b). Figure 4(c) presents the computed Fermi contours for comparison. Six-fold symmetry about the zone center is evident for all cases, which is consistent with a single-domain orientation. Clearly, domain angular averaging, if present, would have an adverse effect on this type of measurement. The closed contours about the zone center $\overline{\Gamma}$ are approximately hexagons. For the 1-ML film, theory shows that there are two such hexagons separated by spin-orbit splitting, which vanishes along the $\overline{\Gamma M}$ direction because of the mirror symmetry as discussed above. The splitting around the hexagon is very small and unresolved in the experiment. For the 2-ML film, the spin-orbit splitting vanishes because of the restoration of inversion symmetry, but the number of bands doubles, resulting in two distorted hexagonal contours separated by interlayer coupling. The Fermi contours around the zone corner $\overline{K}$ point take the form of warped triangles. Two such triangular contours are present for the 1-ML film due to spin-orbit splitting, but the splitting is not resolved in the experiment. For the 2-ML film, the spin-orbit splitting vanishes everywhere in the zone, but the bands double, and so the Fermi map still shows just two closed contours around each $\overline{K}$ point. The splitting caused by interlayer coupling is large and is well resolved in the experiment. The detailed shapes of the contours are somewhat different between theory and experiment. By 4-ML, the measured ARPES Fermi map is already very much bulklike. The only exception is that bulk NbSe$_2$ has an additional small pancake-shaped Fermi contour [13,34,35] around the zone center that is derived from the Se $4p_z$ orbital. Within our experimental resolution of 20 meV, we detect no changes in the band structure and Fermi surface as a function of temperature beyond thermal broadening; evidently, any CDW-induced changes are too small and thus not seen in our experiment.

The interlayer-coupling-induced Fermi contour splitting for the 2-ML film is about twice the spin-orbit splitting for the 1-ML film based on theory (Fig. 4(c)). A visual inspection of the data in Fig. 4(b) suggests that the spin-orbit splitting in the ML should be observable in the experiment based on this comparison, but it is not. It is possible that the spin-orbit splitting for the 1-ML film is actually substantially smaller than the theoretical prediction. In view of the discrepancies between theoretical and experimental dispersion relations at various places in the zone as noted above, this difference is perhaps not surprising. We have tried other calculations including the GGA and PBE approximations, but the results are very similar to the LDA results as shown in the figure. While the substrate is ignored in the calculation, we do not believe that it is a significant factor, as the interfacial bonding is weak and incommensurate. Incidentally, our experimental results for the valence band dispersions and Fermi contours for the 1-ML film agree very well with theoretical results without considering spin-orbit coupling [36].

In summary, we have successively grown high-quality ultrathin NbSe$_2$ films down to 1 ML on Se-passivated GaAs(111)B surface. This substrate, with a large band gap, provides a fairly inert nearly non-interacting base for exploring the electronic structure of the supported NbSe$_2$ films. A further advantage of using this substrate is a well-defined single-domain orientation for the NbSe$_2$ films, a condition essential for detailed ARPES mapping of the band structure and Fermi surface. The results show substantial changes from a 1- to 2-ML films, and smaller changes at higher thicknesses. The observed band structure and Fermi surface are in fairly good overall agreement with first-principles calculations, but some discrepancies in energy eigenvalues and Fermi contour separations are noted. Specifically for the 1-ML film, the theoretically predicted spin-orbit splittings of the Fermi contours are not observed experimentally, suggesting an over-estimation of the splittings by theory. These differences suggest the need for further theoretical investigation. As the film thickness increases, more bands emerge. Considering lifetime broadening of the states, the electronic structure at 4 MLs is already nearly bulklike. This is consistent with prior observation of a rapid convergence of the superconducting transition temperature as a function of film thickness. The transition from 2 MLs to 1 ML is interestingly characterized by the suppression of spatial inversion symmetry and a substantial reduction of the superconducting transition temperature. The lack of inversion symmetry causes the bands to split into spin-orbit components except for special locations in the Brillouin zone where the spin degeneracy is protected by a mirror-plane symmetry. These changes are likely connected to the superconducting properties, but quantitative details will require further theoretical work.

*Acknowledgments* We thank Dr. Fengcheng Wu for helpful discussions and Dr. Jonathan Denlinger for assistance with our experiments at ALS. This work is supported by the U.S. Department of Energy (DOE), Office of Science (OS), Office of Basic Energy Sciences, Division of Materials Science and Engineering, under Grant No. DE-FG02-07ER46383 (T.C.C.). The Advanced Light Source is supported by the Office of Basic Energy Sciences of the U.S. DOE under Contract No. DE-AC02-05CH11231. C.Z.X. received partial support from the ALS Doctoral Fellowship in Residence during the experiment. X.X.W. is supported by the National Science Foundation of China under Grant No. 11204133 and the Fundamental Research Funds for the Central Universities under Grant No. 30917011338. D.F. is supported by the Deutsche Forschungsgemeinschaft (FL 974/1-1).

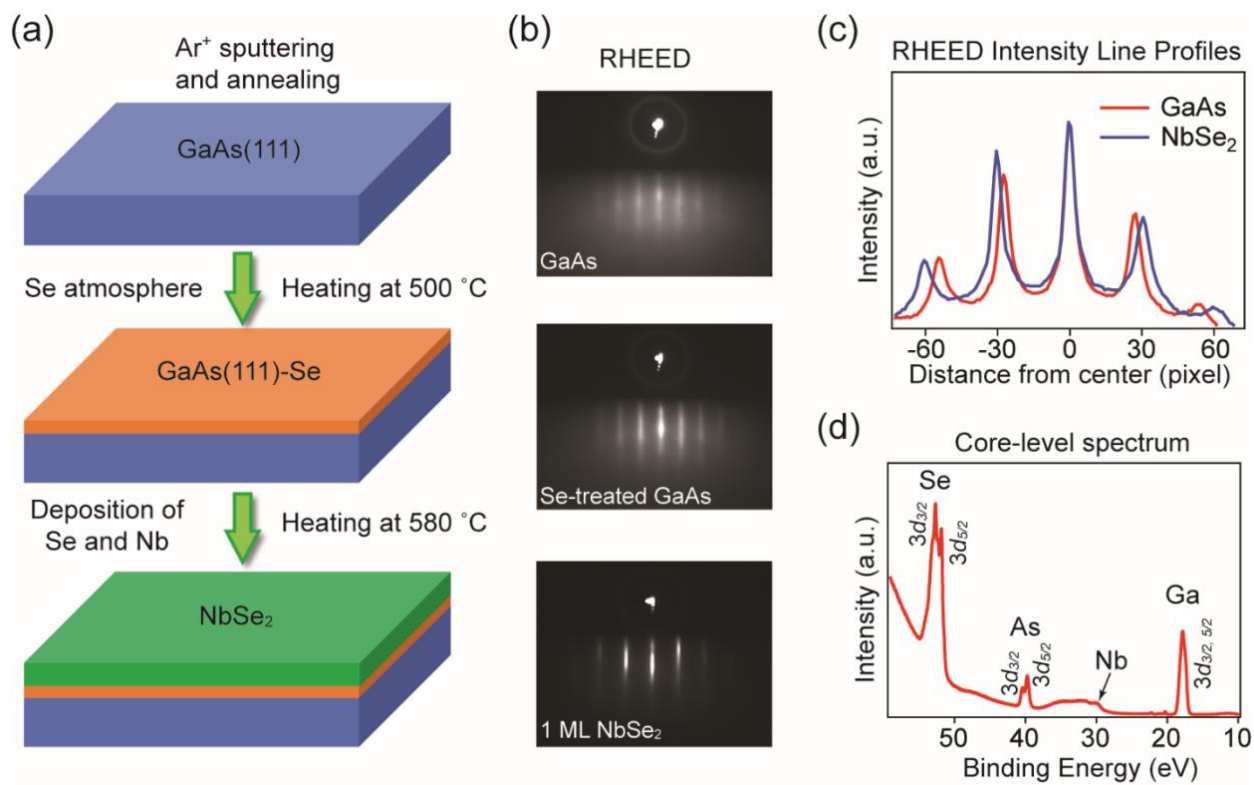

Figure 1. (a) Schematic growth processes for NbSe₂ films on GaAs(111) by MBE. (b) RHEED patterns taken after each step in (a). (c) Intensity profiles for the RHEED patterns of the Se-treated GaAs substrate and the same after growth of a 1-ML NbSe₂ film. (d) Core-level spectrum of 1-ML NbSe₂. The photon energy used was 80 eV.

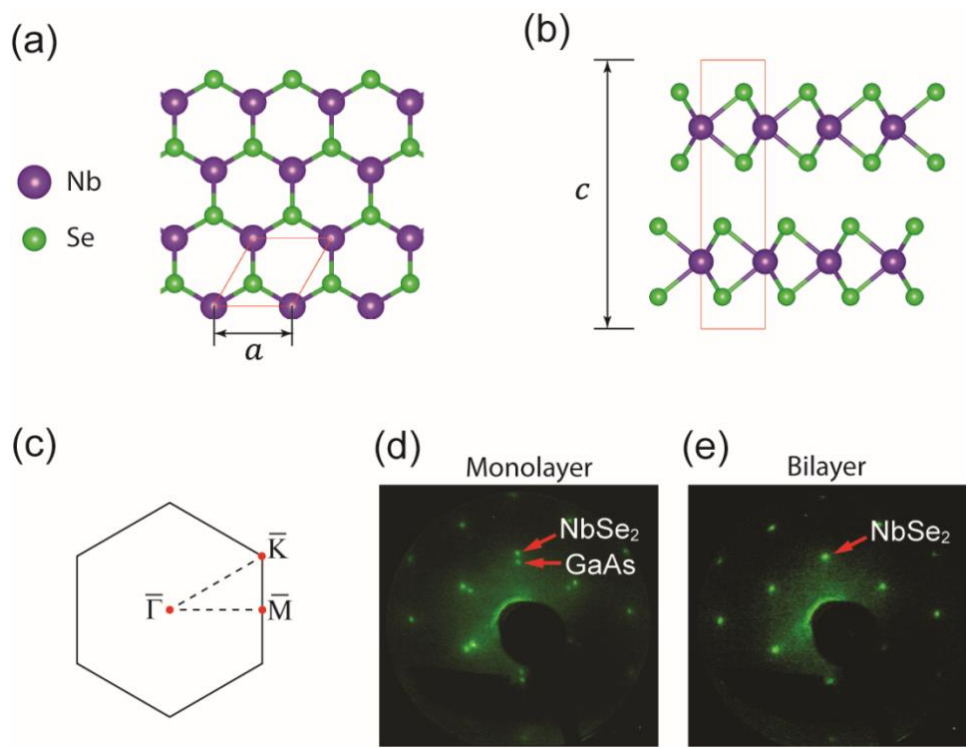

Figure 2. (a) Top view of a schematic atomic structure of NbSe$_2$ ML. (b) Side view of 2-ML NbSe$_2$. The orange rectangle indicates the unit cell of bulk 2H-NbSe$_2$. (c) Brillouin zone of NbSe$_2$ ML. (d) LEED pattern from 1-ML NbSe$_2$. (e) LEED pattern from 2-ML NbSe$_2$. The energy of the LEED beam was 135 eV.

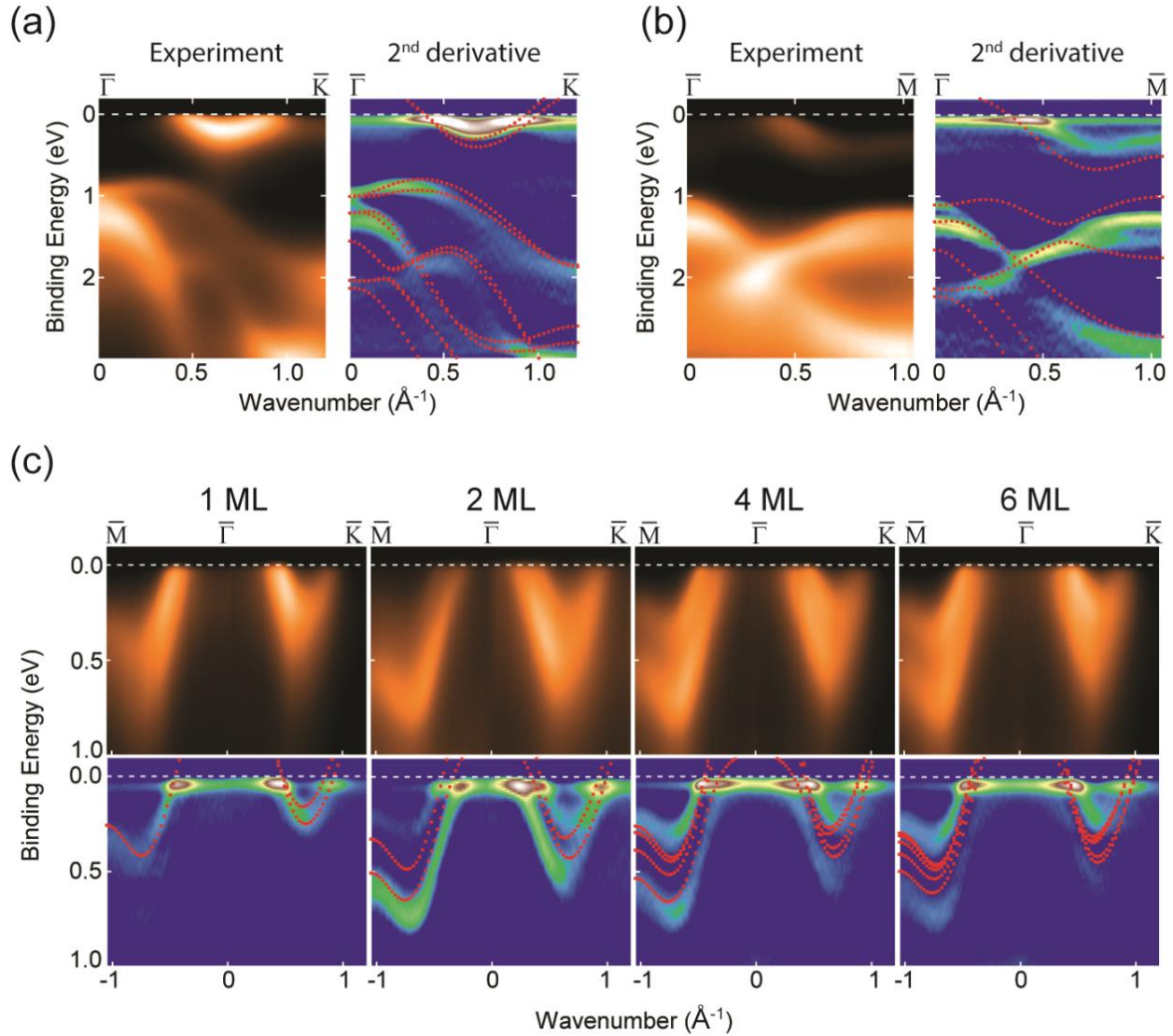

Figure 3. (a) ARPES map from NbSe$_2$ ML along the $\overline{\Gamma}\overline{K}$ direction and the second derivative of the same map. The red dotted curves overlaid on top of the second-derivative map are calculated band dispersion relations. (b) Same as (a) but for the $\overline{\Gamma}\overline{M}$ direction. (c) ARPES maps and the corresponding second-derivative maps for 1-, 2-, 4- and 6-ML NbSe$_2$ films near the Fermi level. Also shown are the theoretical band dispersion relations. The Fermi levels in the calculation were slightly adjusted to better match the experimental results. All ARPES data were taken with 55-eV photons at a sample temperature of ~40 K.

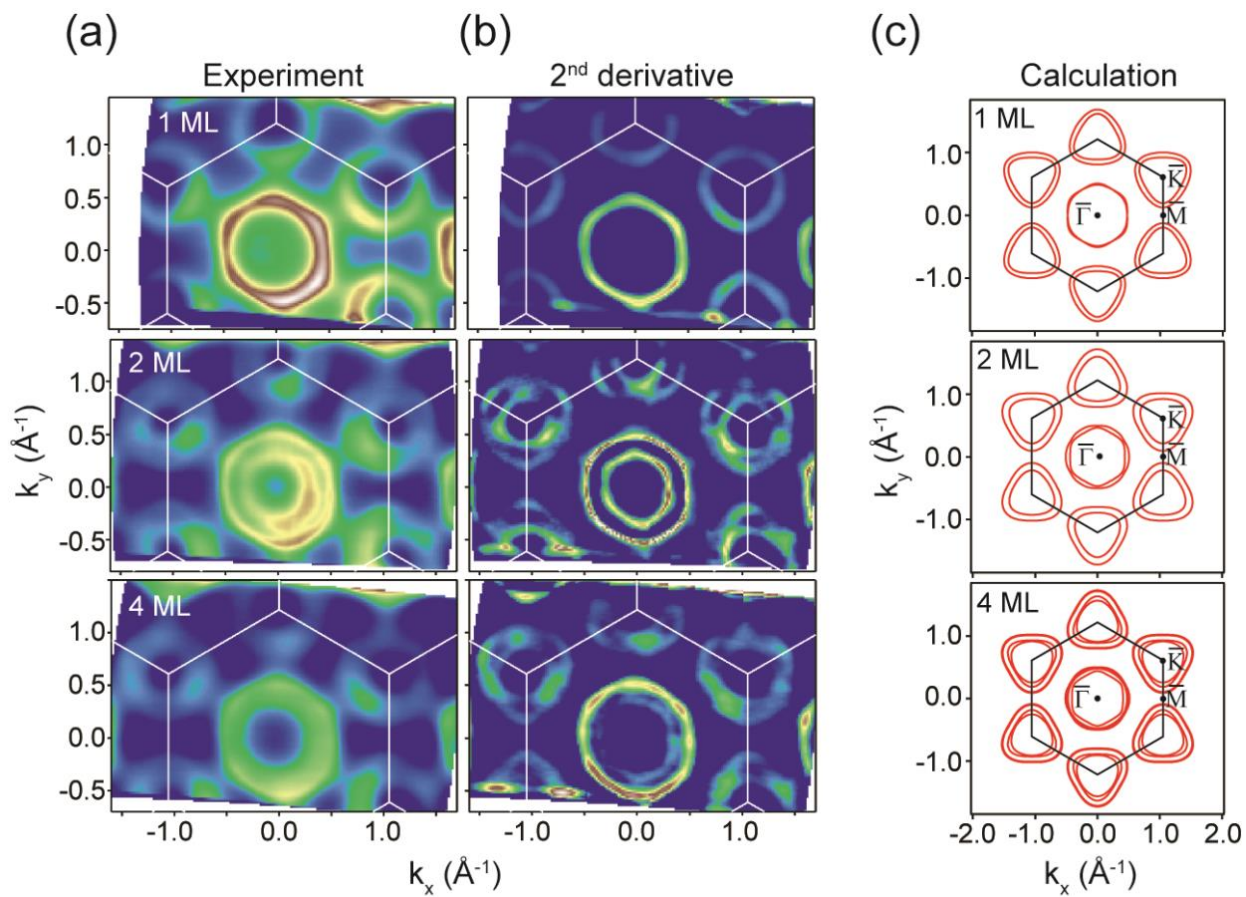

Figure 4. (a) ARPES maps of the Fermi surfaces of 1-, 2-, and 4-ML NbSe$_2$ films. The Brillouin zone is indicated. (b) Corresponding second-derivative maps. (c) Calculated Fermi contours. All ARPES data were taken with 55-eV photons at a sample temperature of ~40 K.